# A New Load-based Thermopower Measurement Setup in the Temperature Range of 5 - 330 K

Tarachand, Monika Saxena, Bodhoday Mukherjee and Gunadhor S. Okram[a)]

*UGC-DAE Consortium for Scientific Research, University Campus, Khandwa Road, Indore-452001 (India).*

[a)]Corresponding author: okram@csr.res.in

**Abstract.** This is the report on new development of an automated precision load based measurement setup for thermoelectric power of different types of samples in the temperature range of 5–330 K. The problems in old spring-based setup have been solved by making load based setup. This new setup takes nearly 4 h for each run and typical error is within 5 %. A high quality calibration has been demonstrated using high purity platinum wires.

## INTRODUCTION

World energy crisis is currently attracting much attentions. Towards this, thermoelectric (TE) materials have been emerged as the potential materials to produce energy, especially from waste heat, silently without releasing any harmful gas. For this however, highly efficient materials are needed and the efficiency of TE materials is directly proportional to the square of Seebeck coefficient (S). On other hand, it is well known that S is (i) highly sensitive for any change in electronic density of states near Fermi level [1], (ii) gives the sign of majority charge carriers, (iii) provides good estimate for charge carrier density and (iv) provides the fundamental information about materials. To enable this, a reliable measurement setup to produce accurate and reproducible data is required. Less time in collecting the data is an added bonus in saving electrical power and liquid helium (LHe) consumption. In this direction, several efforts have already been made [2-7]. Mun *et al.* [4] reported an experimental setup for TEP measurements in temperature range of 2 to 350 K. They measured $\Delta V^{\pm}$, directly from electrical contacts pad made using indium or silver epoxy on top surface of sample [4,7], using an alternating temperature gradient $\Delta T$. However, the difficulty creeps in when the size of the samples is small ($\leq$ 1 mm$^3$): not only that making contacts become very difficult but also with additional small separation-induced error in S [5]. The thickness of the sample also constrained below 0.5 mm to avoid any vertical temperature gradient in the sample that may introduce an extra error in $\Delta V$ [4,7] and hence S. They collected S in both forward and reverse $\Delta T$ that requires 50 to 88 s to establish thermal equilibrium, which can increase at increased thickness ($\geq$ 0.5 mm) [4,7]. On other hand, for making of good thermal and electrical contacts, spring load-based thermoelectric power (TEP) setups are simple, well known and have been the most popular method [2,3,5].

We earlier demonstrated a spring-based load TEP setup [2] in which accurate and reproducible data can be obtained in warming up mode. In this setup, sample loading and de-loading is much quicker and easier than others [4] and it is non-destructive technique. However, the reliability of the equipment or data depends on the strength of spring and its strength depends on material used and its type. Due to the significant temperature difference cycling of from 5 K to 330 K with respect to room temperature (300 K), any spring responds differently from another with associated modified and unreliable data. For this, one has to pay great attention towards the selection of material of the spring in the contact system like contact points of the wiring and complications due to surface oxidation of the copper block that necessitates frequent calibration essential and mandatory to obtain reliable data, in line with also earlier reports [5]. Moreover, in the earlier system [2], cooling was very unreliable and risky with uncertainty of the rubber balloon being exploded when not monitored continuously, yet taking nearly 3 h cooling time with the consumption of about 12 L of LHe altogether for a set of data. In contrast, we report here our successful resolution of these two issues. First, changing of used material and diameters of wires of springs were checked but it was not successful due to fast degradation of springs with repeated cycles of heating and warming. We replace the unreliable spring by a variable loading system free from unreliability of the spring, enabling highly reliable and reproducible data acquisition (Fig. 1). Second, hit and trial methods of introducing the helium gas to cool the sample chamber led us to achieve a record cooling time of just about 1 h and LHe consumption of just about 2 L for one set of data instead of ~3 h cooling time and ~12 L in earlier setup [2].

## DISCRIPTION OF CRYOSTAT AND TEP EXPERIMENTAL SETUP

### I. General purpose cryostat

It is the general purpose vacuum insert (cryostat) made for low temperature thermoelectric power and electrical resistivity measurements [2] and was fabricated using an SS tube (93 cm L X 5 cm ID X 0.05 cm T) brazed with (10 cm L X 5.1 cm ID X 0.2 cm T) oxygen-free highly conducting copper (OFHC) solid rod made as cup at the bottom, a ball valve, and a vacuum port near the neck with the top-end fitted with an appropriate sample holder, where L is length, ID is internal diameter and T is thickness. It can be evacuated down to about $10^{-5}$ mbar vacuum and dips in 60 l liquid helium Dewar for cooling down to 5 K.

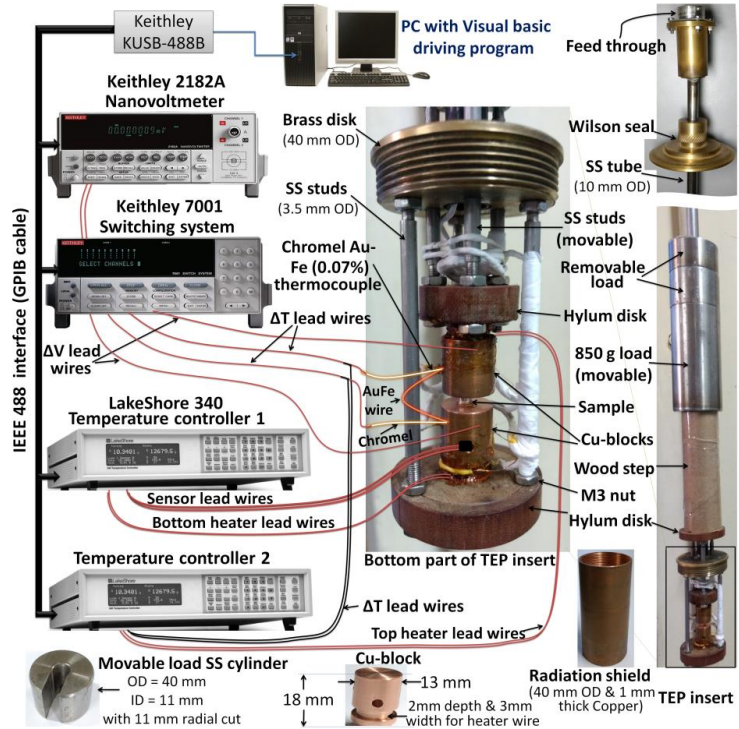

**FIGURE 1.** Schematic diagram of Seebeck coefficient S measurement setup connected to computer-interfaced system with typical photographs of load-based S sample holder wired with nanovoltmeter, switching system and two temperature controllers for T, ΔV and ΔT measurements. Insets: extreme bottom left, a typical detachable load; middle bottom left, copper block; middle right, radiation shield; extreme right top, feed-through upper side and bottom, housing of sample holder. Other details are in legends.

### II. Thermoelectric power measurement sample holder

This is the heart of the system. A freshly cleaned surface sample (2mm x 2mm x 1mm, ideally) is normally mounted between two cylindrical OFHC blocks (13 mm dia x 18 mm L), which are pressurized from a permanently fixed annular stainless steel cylinder of mass ~850 g (40 mm OD and 11 mm ID) loaded from the upper side of the sample holder-housing (Fig. 1, extreme right, bottom inset); OD, outer diameter. Seebeck coefficient (S) is measured in dc differential method of configuration [2,3]. An appropriate hole for inserting one end of a Chromel-Au-Fe (0.07%) differential thermocouple (TC) and a groove on the upper side for heater wire winding are made in upper movable OFHC-block fixed with a hylum disk (Fig. 1, inset, extreme right bottom). The hylum disk is fitted rigidly with three parallel studs having each of them both-ends-threaded and passing freely through three inner holes of a brass disk. The upper sides of the vertically parallel studs are fitted rigidly to a second hylum disk using M3 nuts, on the top of which a load of ~ 850 g is loaded. The lower OFHC-block with appropriate holes (Fig. 1, middle left bottom inset) for other end of the TC and Si-diode sensor (3 mm OD), having a groove at the lower end for winding heater wire are made. It is in turn rigidly fitted to a third hylum disk at the bottom (Fig. 1 and inset, extreme right bottom). The hylum disk is in turn fixed rigidly with a brass disk (40 mm OD x 12 mm thick) on the upper side using three parallel outer studs, with each of them having both-ends-threaded (3.5 mm dia x 90 mm L), and M3 nuts (Fig. 1). Upper OFHC-block can be slid up and down vertically with three parallel SS studs (3.5 OD x 65 mm L) through the three corresponding inner holes of the brass disk. Extra removable loads such as 200 g or 300 g can be loaded as desired (Fig. 1, extreme left, bottom inset).

Once these mechanical arrangements have been made successfully, next comes the wiring that is equally challenging to obtain a reliable S data. In all, lower ends of 12 lead wire terminals of each about 1m length are used for the sample holder. In this, 8 copper wires of 80 μm diameter each are used for sensor, TC and ΔV, and 4 copper wires are of 180 μm diameter each used for heaters are connected to the heaters housed with the two OFHC-blocks. The upper side end terminals of the wires are connected at the inner side female silver-coated pins of the feedthrough (Fig. 1, extreme right top). The wire diameter selections are crucial for appropriate electrical voltage/current to be passed and or restricted heat flow control requirements. Sensor, TC, ΔV and heater terminals are carefully soldered to the corresponding lower end copper wires. For ΔV, the two copper wire terminals are soldered prudently with ohmic contacts directly on the respective clean surfaces of OFHC blocks. For TC, two terminals of chromel wires are soldered to the copper wires and insulated with Teflon tapes.

Fifteen-inch-long insulated manganin heater wire (250 μm diameter) having about 15 Ω each is wound tightly and uniformly over each OHFC groove and soldered the terminals carefully to the copper wires from above sample holder and anchored each insulated joint to a nearby stud as heat sink. Over the manganin heater wire winding, 80 μm diameter 150 cm long insulated copper wire is wrapped tightly and uniformly to enable heat transfer more efficiently and avoid heat leak openly. All the soldered joints are made insulated and fixed rigidly to different studs as and where available nearest. These are essential to obtain consistently good S data. All this constitutes the sample housing (Fig. 1). The sample housing is moreover safely protected by a removable threaded 40 mm OD x 1 mm thick x 90 mm long OFHC cup (Fig. 1, middle, right inset) screwed to the brass disk as a radiation (and or damage-protection) shield. This makes the sample holder handling easy in general while inserting inside the cryostat or removing outside it.

### III. Experimental procedures

The female vacuum feedthrough connector, with all the 12 lead wires for sensor, heaters, TC and ΔV connected, is coupled to the external male coupler to which cables leading to respective temperature controllers, nanovoltmeter, switching system and finally coupled to a personal computer (PC) using respective GPIB cables and a KUSB cable are connected. The PC controls, measures and acquires S data at a desired T from the constituent equipments through a program written in Visual Basic (Fig. 1) when the sample holder is fitted inside the general purpose cryostat that in turn is inserted half its length in a specially designed LHe Dewar strategically after evacuation (~$10^{-5}$ mbar). The sample started showing cooling sign through the Si sensor and temperature controller-1 attached to the lower OFHC-block (Fig. 1). Then, a small amount of (He) exchange gas is introduced into the cryostat so that the sample gets cooled faster. At ~ 100 K, the cryostat is fully inserted or just dipped into the LHe Dewar that leads to cooling the sample T to about 5 K within ~1 h from the starting of inserting the general purpose cryostat.

Temperature controller - 2 controls the temperature (T) difference ΔT between the two OFHC-blocks using the TC. In a typical S (-ΔV/ΔT) measurement at a chosen T, a ΔT is maintained and the corresponding ΔV is measured (Fig. 1), which gives $S_{total}$ = -ΔV/ΔT = $S_s$ (sample) + $S_{ref}$ (OHFC as reference) at T. The program from the PC through temperature controller-1 and measuring system collects initial temperature ($T_i$) from Si sensor attached with colder OFHC-block. The thermal electromotive force ($V_{themf}$) generated across the two junctions of TC is measured through leads connected to 1st channel of switching system using a nanovoltmeter. The $V_{themf}$ is divided by known thermopower of differential TC ($S_{tc}$) to get ΔT (= $V_{themf}/S_{tc}$), in K. After this, the switching system switches into channel 2 to measure ΔV through a nanovolttmeter. Immediately after this, the final temperature ($T_f$) is recorded using temperature controller-1. Then, average T = ($T_i+T_f$)/2 is taken as the sample T; this averaging is required because data will record in ramping mode and hence $T_i$ and $T_f$ are usually different. The ratio of ΔV and ΔT gives $S_{total}$ at T, and S can be extracted from $S_{total} - S_{ref}$, where $S_{ref}$ is the Seebeck coefficient of Cu. Similarly, other data points are recorded at different chosen T and thus the T range desired, say ~5-325 K. Data can be recorded both in cooling and warming cycles; however, more reliable data are normally obtained during warming cycle in which T and ΔT control is easily possible systematically.

## IV. Precautions

The following precautions must be kept in mind.
i) ΔT measured is considered as the actual ΔT at the two ends of the sample as thermal conductivity of OFHC-blocks is significantly very high or infinite. However, practically it is not so. Not only this, thermal transport at the joint between the sample and OFHC also could be other possible complicacies. This is more so, because the OFHC is naturally coated with its oxide even if one tries his best to clean it.
ii) Similar is the case for ΔV as well even though one may assume comfortably that a metal has equipotential surface and hence ΔV on OFHC-blocks are ΔV measured at the end surfaces of the sample. This deviation may be really drastic for old system wherein cleaning oxidized surface of the OFHC is almost impossible as per expected and much deviation from what one expects since S of the surface might range from 10 - 1000 µV/K or even much more. Hence, measurements of S especially for metallic samples (~ 1-10 µV/K) should be extra-careful and recalibrated before each measurement, as far as possible.
iii) Other problems include the aging degradation of any contact points in the whole wiring arrangements, copper blocks, any reference metal or spring, if any, and hence possible deviation of the data from the actual value.

## RESULTS AND DISCUSSION

The most common problem in spring based TEP experimental setup is the contraction of spring at low temperature or not able to restore while warming and causes cracks, extra hump formation or slope change in temperature dependent S data. To resolved this unwanted problem a load based TEP setup has been made with reverse geometry than that reported in [2] as shown in Fig. 1.

To test the reliability of setup the thermopower of Pt wire of diameter 400 µm (99.997 %) is measured with applied load of 850 g using differential dc method with OFHC as reference and warming data under various optimizations are shown in Figure 2 (a). In general, the absolute TEP (S) of a pure Pt (metal) consist of two types of contributions, first $S^e$ arising from the nonequilibrium distribution of the conduction electrons (near 300 K) known as diffusion TEP and second, $S^g$ caused by the interaction between the conduction electrons and the phonon (~60 K) known as phonon drag TEP: $S = S^e + S^g$. In our first attempt for Pt wire of 25 µm diameter (99.995 %), Pt1a, under applied load of 850 g the value of TEP at 300 K is 11.3 µV/K whereas a broad hump observed at low temperature has the maximum value of 8.29 µV/K (Fig. 2(a)). Overall trend looks similar to the TEP of Pt wire [2-9] but shows much greater values than the reported values (~5.7 µV/K at 300 K and ~6.1 µV/K at 64 K) [2-9]. This led us to optimize the arrangements of insert including connections. It is thought to be the main problem that may be from ΔV contacts on OFHC blocks. Therefore, it is remade and TEP is collected for Pt wire (25 µm dia), Ptb, (Fig. 2(a)). The TEP value of Ptb at 300 K is -67 µV/K which is bizarre with absolutely deviated tends. This may be due to improper connection or surface oxidation of OFHC blocks. Then we remade ΔV connections and change the Pt wire and tested for Pt wire of diameter 400 µm (99.997 %), Ptc, under same applied load of 850 g. The TEP data of Ptc looks similar to Ptb with higher value at 300 K. Then, we clean the surface of OFHC blocks using zero size sand paper and tested for Pt wires of diameter 400 µm and 250 µm (both of 99.997%) under applied load of 850 g named as Ptd and Pte, respectively. These Ptd and Pte data also showed similarly higher values with deviated trends.

From above data, it is clear that desired value of TEP cannot be achieved using existing OFHC blocks with less purity and decided to replace both blocks by new ones. The TEP values obtained using new OFHC under applied load of 850 g for Pt wire of 400 µ, Pt2, at 300 K is 6.9 µV/K (Fig. 2(b)) which is greater than that of ref. [2-3,5-8] whereas in low temperature a hump observed centered near 60 K with $S_{max}$ = 4.5 µV/K which is less than that of reported value [2-4,6,7]. These values nearly match with that of ref. [4,9], which suggests requirements for further optimization. Afterward, the ΔV terminals on the OFHC blocks are remade along with realignment of moving studs with load, and measured TEP, Pt3, which shows a favorable increase at ~60 K with an unexpected enhancement at 300 K (Fig. 2(b)). In furtherance, the connections of both ΔV and ΔT (between chromel and Cu lead wires) are remade with care and precision, Pt4, which turned into favorable change (Fig. 2(b)). In this attempt, little favorable enhancement near 65 K increased our interest towards making of proper ΔT junctions. The junctions between Chromel and Cu lead wires have remade using solder wire and tested for same Pt wire, Pt5, with all other arrangements fixed and found below

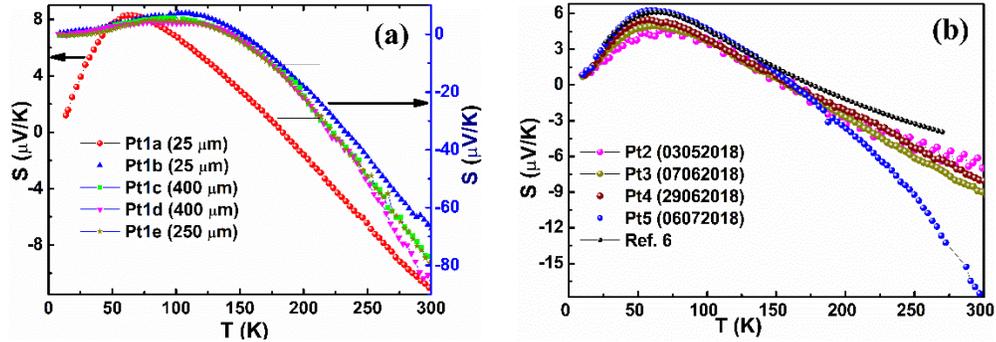

**Figure 2** Temperature dependence of thermoelectric power under applied load of 850g; (a) for pure Pt wires of different diameters as indicated in legend and (b) for Pt wire of diameter 400 μm collected under various optimization conditions, legend indicate dates on which measurements were performed.

150 K TEP data nearly matched with that of Ref. [5-7]. Above this, it shows higher values than previous ones. These attempts indicate making of junction between Chromel wires and Cu-lead wires are very crucial. After this, ΔT junctions are remade using indium wire but due to soft nature of indium wire, junctions are not so strong which can break easily and produce noisy data (SI Fig. S1). We found the best data can be obtained by making ΔT junctions using lead solder only, and optimized one, Pt6, is closely matched with simulated data of ref. [6] (Fig. 3(a) & S2) and well in accordance to other reports [3,5,7,9]. Soni et al. [2] reported cooling thermopower data which is a little deviated from others [5,6] at ≤ 80 K (see Fig. 3(a)), whereas this load-based setup produced data much better and close to ref. [5,6]. In addition, proper thermal anchoring on voltage leads is done on above conditions and found more accurate and smooth S data (Pt7) for Pt wire of 400 μm, S below 150 K almost matches with that in [3,5,6] and above which it is a little deviated from those of [5,6] but close to [2,8,9,10] (Table S1).

For further understanding, we checked it for different Pt samples (Fig. 3(c)). This trend is followed in Pt wire with diameter of 380 μm (Pt8, 99.99%) and 250 μm (Pt9) whereas cylindrical piece (Pt10, 99.97%) sample shows slight deviation (Fig. 3(b)). Cylindrical Pt sample possesses smallest $S^g$ (~60 K) values than others which suggests the requirements of high purity Pt, and perhaps sample shape and size, for getting proper values of phonon drag ($S^g$) maxima.

The thermopower values of Pt7 below 150 K agree well with other reported data in literature [3,5,6] and looks much better than that of [4,9]. Above 150 K, S is a little deviated from that of [5,6] but lies in between those of [5,6] and [2,4,8,9] (Table S1). Therefore, we have fitted our experimental data by extending the equation proposed by Burkov *et al.* [5], which is followed by Abadlia et al. [6] for 25 – 273 K, for thermopower of Pt7 (Fig. 3 (c)); the same equation has some typo in Ref. [7,9]. The modified equation with newly obtained numerical values for S data of Pt wire (Pt7) is

$$S_{Pt}(T) = 0.28234T\left[\exp\left(-\frac{T}{71.579}\right) - 0.05403 + \frac{0.1517}{1+\left(\frac{T}{91.905}\right)^4}\right] - 2.54369 \qquad (1)$$

The difference of experimental thermopower from the function ($\Delta S = S_{exper.} - S_{Pt}(T)$) is ± 0.125 μV/K shown in Fig. 3 (c & d).

We have also tested this setup without load, Pt11, by applying a spring in between moving upper OFHC and fixed brass disk as the bottom part of TEP insert shown in Figure 3(a). At this zero load condition, the value of TEP at below 125 K well accordance with Ref. [6] and also with Pt6, whereas above this, it deviated from Pt6. At high temperature region, the TEP data of Pt11, lies in between that of Ref. [6] and Ref. [4,9]. This system of the TEP measurements can be utilized also for thin film of 1 mm minimum width also by putting vertically in between OFHC blocks using silver paste at edges for electrical contacts [11]. With these optimizations, we have demonstrated a user-friendly (easy to load or de-load the sample), much faster than others [3,4,6,7], accurate and automated with capability of small size sample size (~1 mm$^3$) using less component instruments for low temperature.

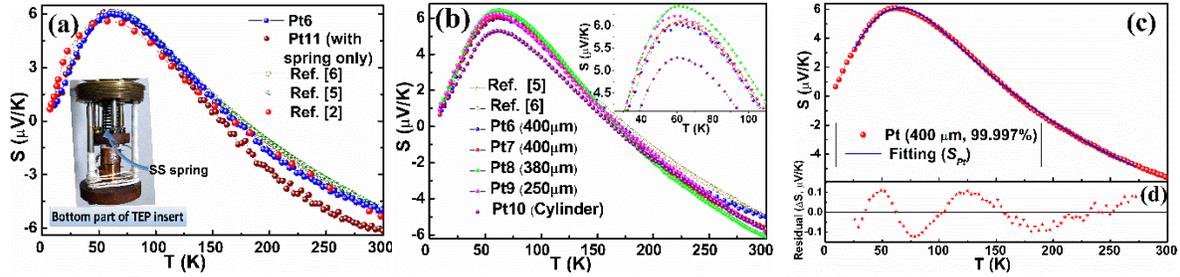

**FIGURE 3** (a) Temperature dependence of thermoelectric power of pure Pt wire (99.997 %) of diameter 400 μm with applied load of 850g and with spring only (Zero load condition) and reported literature as indicated, (b) for different Pt samples as indicated, (c) typical fitting of TEP data of Pt7 (best optimized one) using Eq. (1) and (d) obtained residual TEP (ΔS).

## CONCLUSIONS

We report an experimental setup based on differential dc method for the measurement of the TEP of one sample at a time for various types of samples (single crystal, bar, wire or thin film with less conducting samples like ceramics, metal or semiconductor) in the temperature range from 5 K/ 85K to 330 K using a liquid helium/nitrogen within 4 h. Ramping rate can be increased up to 2 K/min. The maximum uncertainty in TEP is less than 0.4 μV/K depend on sample and temperature range. Measurements of Pt wire (99.999%) are also well-consistent with earlier literature [5,6,7,9].

**Supplementary Material** includes some other data of Seebeck coefficient of Pt wire and others reported earlier for comparison, including in the form a table for greater details.

## ACKNOWLEDGMENTS

Authors are grateful to P. Sarvanan, V. Ganesan and A. K. Sinha from UGC-DAE Consortium for Scientific Research (CSR), Indore for providing cryogens for experiment, and for their encouragements, respectively. Thanks are also due to R. R. Philip (Union Christian College, Aluva, Kerala, India), R. Rawat and S. R. Potdar (UGC-DAE CSR) for providing platinum specimens.

## REFERENCES

1.  J. P. Heremans, V. Jovovic, E. S. Toberer, A. Saramat, K. Kurosaki, A. Charoenphakdee, S. Yamanaka and G. J. Snyder, *Science* **321**, 554-557 (2008).
2.  A. Soni and G. S. Okram, *Rev. Sci. Instrum.* **79**, 125103 (2008).
3.  L. S. S. Chandra, A. Lakhani, D. Jain, S. Pandya, P. N. Vishwakarma, M. Gangrade and V. Ganesana, *Rev. Sci. Instrum.* **79**, 103907 (2008).
4.  E. Mun, S. L. Bud'ko, M. S Torikachvili and P. C. Canfield, *Meas. Sci. Technol.* **21**, 055104 (2010).
5.  A. T. Burkov, A. Heinrich, P. P. Konstantinov, T. Nakama and K. Yagasaki, *Meas. Sci. Technol.* **12**, 264-272 (2001).
6.  L. Abadlia, F. Gasser, K. Khalouk, M. Mayoufi and J. G. Gasser, *Rev. Sci. Instrum.* **85**, 095121 (2014).
7.  T. S. Tripathi, M. Bala, and K. Asokan, *Rev. Sci. Instrum.* **85**, 085115 (2014).
8.  R. P. Huesener, *Phys. Rev.* **140**, A1834-1844 (1965).
9.  T. S. Tripathi and M. Karppinen, *Meas. Sci. Technol.* **30**, 025602 (2019).
10. N. Cusack and P. Kendall, Proc. Phys. Soc. **72**, 898-901 (1958).
11. B. G. Nair, G. S Okram, J. Naduvath, T. Shripathi, A. Fatima, T. Patel, R. Jacob, K. Keerthi. S. K. Remillard, R. R. Philip, *J. Mater. Chem. C* **2**, 6765-6772 (2014).